\documentclass[showpacs,preprintnumbers,amsmath,amssymb,superscriptaddress,nofootinbib]{revtex4}
\usepackage{graphicx}
\usepackage{epsfig}
\begin{document}
\topmargin -2cm
\title{Dissipative generalized Chaplygin gas as phantom dark energy}
\author{Norman Cruz}
\altaffiliation{ncruz@lauca.usach.cl} \affiliation{Departamento de
F\'\i sica, Facultad de Ciencia, Universidad de Santiago, Casilla
307, Santiago, Chile.\\}
\author{Samuel Lepe}
\altaffiliation{slepe@ucv.cl} \affiliation{Instituto de F\'\i
sica, Facultad de Ciencias B\'asicas y Matem\'aticas, Pontificia
Universidad Cat\'olica de Valpara\'\i so, Avenida Brasil 2950,
 Valpara\'\i so, Chile.}
\author{Francisco Pe\~na}
\altaffiliation{fcampos@ufro.cl}
\affiliation{Departamento de Ciencias F\'\i sicas, Facultad de
Ingenier\'\i a, Ciencias y Administraci\'on, Universidad de la
Frontera, Avda. Francisco Salazar 01145, Casilla 54-D, Temuco,
Chile.\\}

\begin{abstract}
The generalized Chaplygin gas, characterized by the equation of
state $p=-\mathcal{A}/\rho^{\alpha}$,  has been considered as a
model for dark energy due its dark-energy-like evolution at late
times. When dissipative processes are taken account, within the
framework of the standard Eckart theory of relativistic
irreversible thermodynamics,  cosmological analytical solutions
are found. Using the truncated causal version of the
Israel-Stewart formalism, a suitable model was constructed which
cross the $w=-1$ barrier. The future-singularities encounter in
both approaches are of a new type, not included in the
classification presented by S. Nojiri, S. D. Odintsov and S.
Tsujikawa, Phys. Rev. D71, 063004 (2005).

\vspace{0.5cm}
\end{abstract}
\maketitle
\section{Introduction}

In the framework of general relativity, the acceleration in the
expansion of the universe during recent cosmological times, first
indicated by Supernovae observations~\cite{Perlmutter} and also
supported by the astrophysical data obtained from WMAP indicates
the existence of a exotic fluid with negative pressure, which
constitutes about the $70$ per cent of the total energy of the
universe. To know the nature and the behavior of this dark energy
is one of the great challenges of cosmology and, indeed, of
fundamental physics.

Within the different candidates to play the role of the dark
energy, the generalized Chaplygin gas (GCG), in which the
pressure $p$ and the energy density $\rho$ are related by
\begin{eqnarray}\label{CGC}
p = -\mathcal{A}/\rho^{\alpha},
\end{eqnarray}
where $\mathcal{A}$ and $\alpha$ are constants, has emerged as a
possible unification of dark matter and dark energy, since its
cosmological evolution is similar to an initial dust like matter
and a cosmological constant for late times. This idea was first
proposed in~\cite{Bento}. Using supernova observations this model
has showed to be consistent for any value of $\alpha$ in the range
$0\leq \alpha \leq 1$~\cite{Makler1}.

Nevertheless, L. Amendola \textit{et al} ~\cite{Amendola} found
that the latest CMB anisotropy data seems favoured GCG as a dark
energy model instead of a candidate for a unified dark model. The
parameter $\alpha$ is rather severely constrained, i.e., $0\leq
\alpha< 0.2$ at the 95$\%$ confidence level. Others works, such
as ~\cite{Carturan} tends to confirm that adiabatic Chaplygin gas
is ruled out, as unified model, unless the parameter $\alpha$ be
very close to zero. A Chaplygin quartessence model with vanishing
pressure perturbation, the called \textit{Silent Chaplygin}, has
shown to be consistent with CMB data~\cite{AmenWaga} .

Since we are interested here only in the late time acceleration
and the phantom behavior, we will considered a GCG as a model for
dark energy.  Specifically, our principal aim is to investigate
the possibilities that GCG models could realize the phantom divide
$\omega =-1$ crossing phenomenon. Previous works have indicated
this possibility. For example, using an extension of the Chaplygin
gas model, where $\mathcal{A}$ was taken as a function of the
scale factor, $a$, of the form $\mathcal{A_{0}}a^{-m}$ (
$\mathcal{A_{0}}$ is a constant and $m>0$), Meng \textit{et al}
~\cite{Meng} showed that exist some cases which can realize state
parameter $\omega =-1$ crossing.

Other approaches have considered dissipative effects in GCG
models, using the framework of the non causal Eckart
theory~\cite{Eckart}. In~\cite{Zhai} a viscous GCG was
investigated from the point of view of the dynamical analysis,
assuming that there is bulk viscosity in the linear barotropic
fluid and GCG. It is found that the equation of state of GCG can
cross the boundary $\omega =-1$.

It is interesting to note that the a GCG itself may behave like a
fluid with viscosity, in the context of the Eckart formalism. J.
C. Fabris {\it et al}~\cite{Fabris} found an equivalence between
the GCG and a dust like fluid with an special election of both
parameter $m$ and $\alpha$, where $m$ is related with the
viscosity coefficient throughout the relation $\xi (\rho)=\beta
\rho^{m}$.

It is well known that the Eckart'approach is a non causal theory.
The full causal theory developed by Israel and
Stewart~\cite{Israel} leads to stable behavior under a wide range
of conditions and transient phenomena on the scale of the mean
free path/time~\cite{Martens}. In this theory the character of
the evolution equation is very complicated, nevertheless we were
able to find solutions in the truncated case for dissipative GCG
cosmological models.

In the present paper we have investigated the behavior of a flat
universe filled with a viscous GCG, using first the non causal
approach of Eckart and then the truncated version of the
Israel-Stewart formalism. In the context of the Eckart formalism,
we show analytical solutions, which presents future-singularities,
which were found restricting the range of the parameters $m$ and
$\alpha$.

Using a truncated version of the Israel-Stewart formalism we were
able to find a suitable model that present future-singularities,
but with an equation of state corresponding to quintessence. This
model has a relaxation time compatible with the expansion of the
universe in a regime near to the thermodynamic equilibrium.

The organization of the paper is as follows: in Section II we
present the field equations for a flat FRW universe filled with a
viscous GCG within the framework of the Eckart theory.
Cosmological solutions are obtained solving a non linear
differential equation for the Hubble parameter. Exact solutions
are obtained in terms of LerchPhi function for  $m$ and $\alpha$
satisfying the constraint given by $m=-(\alpha +1/2)$. In Section
III we solve the equations of the truncated version of the
Israel-Stewart formalism, using an Ansatz for the Hubble
parameter. In section IV we characterize the future-singularities
obtained. In section V we summarize our results.

\section{The Generalized Chaplygin gas and the Eckart theory}
The FRW metric for an homogeneous and isotropic flat universe is
given by
\begin{eqnarray}\label{ds}
ds^2=-dt^2+a(t)^2\left(dx^2+dy^2 + dz^2 \right),
\end{eqnarray}
where $a(t)$ is the scale factor and $t$ represents the cosmic
time. In the following we use the units $8\pi G=1$. The field
equations in the presence of bulk viscous stresses are
\begin{eqnarray}\label{tt}
\left(\frac{\dot{a}}{a}\right)^2=H^2= \frac{\rho}{3},
\end{eqnarray}
\begin{eqnarray}\label{rr}
\frac{\ddot{a}}{a}=\dot{H}+H^2=-\frac{1}{6} \left(\rho + 3 (P +\Pi
) \right),
\end{eqnarray}
where $\Pi$ is the bulk dissipative pressure. The conservation
equation is given by
\begin{eqnarray}\label{ConsEq}
\dot{\rho}+ 3 H (\rho+p+\Pi)=0.
\end{eqnarray}
Assuming that the dark component is GCG, the equation of state
can be rewritten in the form
\begin{eqnarray}
\label{eqstatechap} p\left( \rho \right) =-M^{4\left( 1+\alpha
\right) }\rho ^{-\alpha },
\end{eqnarray}
where $M$ has dimension of mass and $\alpha$ is a constant. The
original Chaplygin gas correspond to the election $\alpha =1$. We
note that $\alpha=-1$ yields the state equation corresponding to
a cosmological constant.

If we put the Eq.~(\ref{eqstatechap}) in the barotropic form
$p(\rho)=\omega \rho$, the parameter $\omega$ take the following
form in terms of the Hubble parameter
\begin{eqnarray}
\label{omegaH} \omega \left( H,\alpha \right)
=-\left(\frac{M^{4}}{3H^{2}}\right)^{1+\alpha}.
\end{eqnarray}
From Eqs.~(\ref{tt}) to~(\ref{omegaH}) is direct to obtain a
single evolution equation for $H$:
\begin{eqnarray}\label{Hpunto}
2\dot{H}+ 3 \left( 1+\omega (H,\alpha) \right)H^{2}= -\Pi.
\end{eqnarray}

In the first order thermodynamic theory of Eckart~\cite{Eckart}
$\Pi$ is given by
\begin{eqnarray}\label{xi}
\Pi=-3 H \xi,
\end{eqnarray}
and, in order to obtain solutions of the equation~(\ref{Hpunto}),
we will assume that the viscosity has a power-law dependence upon
the density
\begin{eqnarray}\label{xilaw}
\xi=\beta \rho^{m}, \,\,\,\,\beta \geq 0.
\end{eqnarray}
where $\beta$ and $m$ are constant parameters. Using
Eq.~(\ref{omegaH}),~(\ref{xilaw}) and~(\ref{xi}) in
Eq.~(\ref{Hpunto}), we obtain the non linear equation leading the
evolution of the Hubble parameter
\begin{eqnarray}\label{Hgeneral}
2\overset{\cdot }{H}+3\left\{ H^{2}-\left[ \left(
\frac{M^{4}}{3}\right) ^{1+\alpha }H^{-2\alpha }+3^{m}\beta
H^{2m+1}\right] \right\} =0.
\end{eqnarray}


Despite the complexity of this equation it is possible to obtain
analytic solutions for special cases in which the parameters
$\alpha$, $\beta$ and $m$ are constrained. In the following
subsections these solutions are explicitly showed.

\subsection{Solutions with  $m=-(\alpha +1/2)$}

If in Eq.~(\ref{Hgeneral}) we choose the exponents of $H$ to be
equal, i.e., $-2\alpha =2m+1$ we obtain
\begin{eqnarray}\label{Hgen}
2\dot{H}+3H^{2}-\beta (\alpha)H^{-2\alpha }=0,
\end{eqnarray}
where $\beta (\alpha)$ is defined by
\begin{eqnarray}\label{betadealfa}
\beta (\alpha) \equiv 3^{1/2 -\alpha}\left[
\beta+\frac{1}{\sqrt{3}} M^{4(1+\alpha)}\right].
\end{eqnarray}
Integrating Eq.~(\ref{Hgen}) we obtain an implicit relation for
$H$ as a function of the cosmic time $t$
\begin{eqnarray}\label{Hgenintegrated}
t&=&\frac{H^{-1}}{3(1+\alpha)}LerchPhi
\left(\frac{\beta(\alpha)}{3}H^{-2(1+\alpha)},1,\frac{1}{2(1+\alpha)}\right),
\end{eqnarray}
where the LerchPhi function is defined as follows
\begin{eqnarray}\label{LerchPhi}
LerchPhi (z,a,\nu ) =\sum ^{\infty}_{n=0}\frac{z^{n}}{(n
+\nu)^{a}}.
\end{eqnarray}
It was mentioned in~\cite{Fabris} that a GCG (without viscosity)
is equivalent to a dust with viscosity, if the parameters
$\alpha$ and $m$ satisfy the constraint $m=-(\alpha +1/2)$. In
our case this constraint was imposed in order to obtain analytic
solutions of a GCG with viscosity.


\subsection{Solutions with future-singularities in the non causal approach}

It is easy to show graphically that above solutions, for positive
values of $\alpha$, represents universes with
future-singularities. Before to show an analytical solution of
Eq.~(\ref{Hgeneral}), let us discuss some interesting results
related to a pure GCG without dissipation. From
Eqs.~(\ref{eqstatechap}) and~(\ref{ConsEq}), we obtain the
evolution of the energy density $\rho$ in terms of the scale
factor $a$, or equivalently in terms of the redshift $z=a_{0}/a
-1$. The expression for $\rho (z)$ is given by (for $\alpha \neq
-1$)
\begin{eqnarray}
\label{rhoz} \rho =M^{4}\left[ 1+g\left( 1+z\right) ^{3\left(
1+\alpha \right) }\right] ^{1/\left( 1+\alpha \right) },
\end{eqnarray}
where $g$ is a parameter that can be fitted by observational data.

Using the above equation in~(\ref{eqstatechap}) we can obtain an
expression for $\omega$ in terms of the redshift
\begin{eqnarray}
\label{megaz} \omega \left( z,\alpha \right) =-\left(1+g\left(
1+z\right)^{3\left( 1+\alpha \right)} \right)^{-1}.
\end{eqnarray}

From Eq.~(\ref{megaz}) is direct to see that for negative $z$
(future cosmic time), we can obtain $\omega =-1$. The phantom
case can be obtained if $g<0$. A. Sen and R. J.
Scherrer~\cite{Sen} explore the possible scenarios when the usual
constraint on the parameters $\alpha$ and $g$ are relaxed. They
consider the possibility of take $g<0$ and $\alpha <-1$. Two
cases of phantom dark energy are obtained. One of them is a late
phantom GCG model which gives $\omega < -1$ with $\omega$
decreasing with time. A future singularity occurs at finite value
of the cosmic time.

In our scheme it is possible to obtain phantom evolution in the
special case, which is straightforward to integrate, when
$m=1/2$. The bulk viscosity coefficient $\xi$ takes the form
\begin{eqnarray}\label{xiunmedio}
\xi =\beta \rho ^{1/2}=\sqrt{3}\beta H.
\end{eqnarray}
Introducing Eq.~(\ref{xiunmedio}) in Eq.~(\ref{Hpunto}), we obtain
\begin{eqnarray}\label{Hunmedio}
2\dot{H}+3\left( 1-\sqrt{3}\beta \right) H^{2}=-3\omega \left(
H\right) H^{2},
\end{eqnarray}
which with the election $\beta =1/\sqrt{3}$, the above equation
has the solution
\begin{eqnarray}\label{Hsol}
H\left( t\right) =H_{0}\left[ 1+\frac{3}{2}\left( 2\alpha
+1\right) \left( \frac{M^{4}}{3H_{0}^{2}}\right) ^{1+\alpha
}\left( t-t_{0}\right) \right] ^{1/\left( 2\alpha +1\right) }.
\end{eqnarray}
Assuming that $2\alpha +1 =-\delta < 0$, the
solution~(\ref{Hsol}) can be rewritten as
\begin{eqnarray}\label{Hsolu}
H\left( t\right) =H_{0}\left[ 1-\frac{\left( t-t_{0}\right)
}{t_{s}}\right] ^{-1/\delta },
\end{eqnarray}
obtaining,  from Eq.~(\ref{rr}), the explicit expression for the
energy density ,
\begin{eqnarray}\label{Hsolu}
\rho\left( t\right) =3H_{0}^{2}\left[ 1-\frac{\left(
t-t_{0}\right) }{t_{s}}\right] ^{-2/\delta },
\end{eqnarray}
which blows up at the time $t_{s}$, given by
\begin{eqnarray}
t_{s}=\frac{2}{3\delta }\left( \frac{3H_{0}^{2}}{M^{4}}\right)
^{\left( 1-\delta \right) /2}.
\end{eqnarray}
The solution given in Eq.~(\ref{Hsolu}) implies that
$\dot{H}+H^{2}>0$, i.e., the acceleration is positive. An
expression for $a(t)$ can be easily obtained integrating
Eq.~(\ref{Hsolu})
\begin{eqnarray}
a\left( t\right) =a_{0}\exp \left\{ \frac{\delta }{\delta
+1}H_{0}t_{s} \left[ \left( 1-\frac{\left( t-t_{0}\right)
}{t_{s}}\right) ^{-\left( 1+\delta \right) /\delta }-1\right]
\right\},
\end{eqnarray}
which at $t-t_{0}\rightarrow t_{s}$ blows up, representing a
solution with a future-singularity. In brief, a GCG with bulk
viscosity contains big rip solutions if $\alpha <-1/2$ and if the
bulk viscosity coefficient takes the specific form $\xi
=\frac{1}{\sqrt{3}}\rho ^{1/2}$.

It is interesting to compare the above results with those
obtained in~\cite{Sen}, where the model with a future singularity
occurs when $g<0$ and $\alpha <-1$. Nevertheless, it is easy to
show from Eq.~(\ref{megaz}) that if $g<0$ it is possible to find,
in the past or in the future evolution of the universe (depending
on the value of $g$) that the fluid behaves as a stiff matter.
This result is not physically plausible as we know from the
cosmological observations that ruled out stiff matter, which
would be of the importance only in the very early universe (see,
for example~\cite{Banks}). In our case, a CGC fluid with bulk
viscosity, the solution with future singularity is obtained for
$g>0$ and $\alpha < -1/2$.

Note the exponential behavior of the scale factor with time. In
the case where dark energy component obey the state equation
$p=(\gamma-1) \rho$ and $\xi \sim \rho^{1/2}$ there are also big
rip solutions but with a power-law expansion for the scale factor
of the form
\begin{eqnarray}\label{scalefactor}
a(t)=a_{0}\left(1-\frac{t-t_{0}}{t_{s}}\right)^{\frac{2}{3(\gamma
-\sqrt{3}\beta)}},
\end{eqnarray}
where $a_{0}=a(t=t_{0})$. This solution was found by
Barrow~\cite{Barrow} and if the following constrains on the
parameters $\beta$ and $\gamma$ are imposed~\cite{Cruz}
\begin{eqnarray}\label{constrain}
\sqrt{3}\beta > \gamma,
\end{eqnarray}
the scale factor blows up to infinity at a finite time
$t_{s}>t_{0}$, given by
\begin{eqnarray}\label{tbig}
t_{s}=\frac{2}{3(\sqrt{3}\beta -\gamma)}H_{0}^{-1}.
\end{eqnarray}.

\section{A truncated version of the Israel-Stewart formalism}

In this approach we consider an imperfect fluid whose dissipative
bulk pressure, $\Pi$, obeys the causal transport
Eq.~\cite{Maartens}
\begin{eqnarray}\label{xitau}
\Pi +\tau \dot{\Pi}=-3 H \xi,
\end{eqnarray}
where $\tau$ is the bulk relaxation time and is the crucial
thermodynamic parameter in the Israel-Stewart formalism that
ensures causality.

In order to obtain an equation for $\Pi$ in terms of Hubble
parameter $H$, we derive Eq.~(\ref{Hpunto}) introducing then
Eq.~(\ref{xitau}). This yields
\begin{eqnarray}\label{Pipunto}
\Pi (H)= \tau \left( 2\ddot{H}+ 3 \dot{\omega}(H,\alpha)
H^{2}+6(1+ \omega (H,\alpha))H\dot{H} \right)-3\xi H.
\end{eqnarray}
The parameter $\tau$ is related to the energy density throughout
the equation
\begin{eqnarray}\label{taulaw}
\tau=\frac{\xi}{\rho}=\beta \rho^{m-1},
\end{eqnarray}
where $\xi$ is given by Eq.~(\ref{xilaw}). In~\cite{Cruz} big rip
solutions were found within the framework of the full causal
theory of Israel-Stewart, for universes filled a dark component
obeying the state equation $p=(\gamma -1)\rho$. For this case,
using the Ansatz $H=A(t_{s}-t)^{-1}$, where $A\equiv H_{0}t_{s}$
and $t_{s}$ is the time when the future-singularity occurs, we
were able to find the exact expression for $A$.

\subsection{Solutions with future-singularities in the causal approach}

In the following we will generalize the expression for the Ansatz
thinking in the more complex structure of the equation of state
corresponding to a GCG. Our election is
\begin{eqnarray}\label{AnsatzH}
H=A \Delta ^{-q},
\end{eqnarray}
where $A=H_{0}(t_{s}-t_{0})^{q}$, $\Delta \equiv t_{s}-t$ and $q$
is a positive constant, which is determined by consistence of the
equations. Using the above Ansatz and Eqs.~(\ref{tt}),
~(\ref{omegaH}) and~(\ref{xilaw}), in Eq.~(\ref{Pipunto}), we
obtain
\begin{eqnarray}\label{abcdpunto}
\Pi (\Delta)= -3^{m}\beta \Delta ^{-2q(m+1/2)} \left(
a(m)-b(m,q)\Delta ^{q-1}-c(m,q)\Delta ^{2(q-1)}-\alpha
d(m,q;\alpha)\Delta ^{q(3+2\alpha)-1} \right),
\end{eqnarray}
where the coefficients $a(m)$ and $b(m,q)$ are given by the
following expressions
\begin{eqnarray}\label{ab}
a(m)=3A^{2(m+1/2)}, \,\,\,\,\,\,\,\,\,\,\,b(m,q)= 2qA^{2m},
\end{eqnarray}
and $c(m,q)$ and $d(m,q;\alpha)$ by
\begin{eqnarray}\label{ab}
c(m,q)= \frac{2}{3}q(q+1)A^{2(m-1/2)},\,\,\,\,\,\,\,\,
d(m,q;\alpha)=2q \left(\frac{M^{4}}{3}\right)^{1+\alpha}
A^{2(m-1-\alpha)}.
\end{eqnarray}
{\bf Condition $\Pi<0$}. Since from thermodynamics arguments (as
requiered from the second law), $\Pi<0$, we will explore the
possible scenarios which are obtained depending on the values of
the parameters $q,m,$ and $\alpha$.

{\bf Case $q=1$}. Let us investigate  the special case with $q=1$,
where the factors with the time dependence $\Delta ^{q-1}$ in
Eq.~(\ref{abcdpunto}), becomes constant. In this situation we
obtain constraint for $A$ which are independent of the parameter
$m$. In brief, we have the following possibilities in order to
ensure $\Pi <0$:
\begin{eqnarray}\label{A}
A>\frac{1+\sqrt{5}}{3}, \,\,\,\,\,\,\,\,\,\,\,\alpha=0,
\end{eqnarray}
and
\begin{eqnarray}\label{A1}
A\geq \frac{1+\sqrt{5}}{3}, \,\,\,\,\,\,\,\,\,\,\,\alpha<0.
\end{eqnarray}
In the first case it is straightforward to see that the time
$t_{s}$ has a lower limit given by
\begin{eqnarray}\label{tb}
t_{s} > \frac{1+\sqrt{5}}{3}H_{0}^{-1}.
\end{eqnarray}

{\bf Case $q>1$}. Choosing for simplicity $\alpha =0$, the sign
of Eq.~(\ref{abcdpunto}) becomes dependent of the sign of the
factor $ a(m)-b(m,q)\Delta ^{q-1}-c(m,q)\Delta ^{2(q-1)}$. Since
the powers of the $\Delta$ functions are both positives, they are
decreasing functions of the cosmic time. This implies that if we
demand the above factor to be positive at $t=0$ it will positive
for the future evolution. Solving the constraint
\begin{eqnarray}\label{abcdto}
\left( a(m)-b(m,q)\Delta ^{q-1}-c(m,q)\Delta ^{2(q-1)}
\right)|_{t=0}>0,
\end{eqnarray}
and since $a(m),b(m,q),c(m,q)$ are functions of
$A=H_{0}(t_{s}-t_{0})^{q}$, at $t=0$ Eq.~(\ref{abcdto}) gives a
constraint for $t_{s}$ in the form of $\mathcal{P}(t_{s})>0$,
where $\mathcal{P}(t_{s})$ is a polinomial of the form
\begin{eqnarray}\label{polinom}
\beta _{1}t_{s}^{q}-\beta_{2}t_{s}^{q-1}-\beta_{3}t_{s}^{-q},
\end{eqnarray}
which can be solved for $t_{s}$.

{\bf Case $0<q<1$}. For this case $\Pi <0$ if $\alpha <0$

{\bf Condition $|\Pi|<<|p|$}. We have just found the constraints
on the parameters $\alpha$, depending of which values are given
for $q$, in order to obtain a negative dissipative pressure. Note
that these constraints are independent of $m$. Constraints on $m$
appears when we consider that the Israel-Stewart theory is
derived under the assumption that the thermodynamical state of
the fluid is close to equilibrium, which means that the
nonequilibrium bulk viscous pressure should be small when
compared to the local equilibrium pressure that is $|\Pi|<<|p|$.
Using Eqs.~(\ref{abcdpunto}), (\ref{eqstatechap}) and
(\ref{omegaH}) the above condition takes the form
\begin{eqnarray}\label{Pionp}
\frac{|\Pi (\Delta)|}{|p(\Delta)|}= 3^{m-1} \beta \left(
\frac{M^{4}}{3}\right)^{-(1+\alpha)} A^{2\alpha}\Delta
^{-2q(m+1/2+\alpha)}\times \nonumber \\
\left( a(m)-b(m,q)\Delta ^{q-1}-c(m,q)\Delta ^{2(q-1)} -\alpha
d(m,q;\alpha)\Delta ^{q(3+2\alpha)-1} \right)<<1.
\end{eqnarray}
It is straightforward to check out that both conditions, $\Pi <0$
and $|\Pi|<<|p|$ are satisfied in the following cases:

{\bf Case $q=1$}

\begin{eqnarray}\label{A2}
\alpha=0,\,\,\,\,\,\,\,\,\,\,\,\ m+\frac{1}{2}<0,
\end{eqnarray}
and
\begin{eqnarray}\label{A3}
A= \frac{1+\sqrt{5}}{3},
\,\,\,\,\,\,\,\,\,\,\,\alpha<0\,\,\,\,\,\,\,\,\,\,\,\
m<\frac{1}{2}.
\end{eqnarray}

{\bf Case $q>1$}

\begin{eqnarray}\label{A4}
\alpha=0,\,\,\,\,\,\,\,\,\,\,\,\ m+\frac{1}{2}<0.
\end{eqnarray}

We finally found the value of the parameter $\beta$, constraining
the relaxation time $\tau$ to be lower than the Hubble time, i.e.,
$\tau <H^{-1}$. The expression for $\tau$ is given by
\begin{eqnarray}\label{tauH}
\tau=\frac{\xi}{\rho}= 3^{m-1}\beta A^{2(m-1/2)}\Delta
^{-2q(m-1/2)} H^{-1}.
\end{eqnarray}
Since the relation $\tau <H^{-1}$ must be satisfied for all
times, at $t=0$ we obtain and upper limit for the parameter
$\beta$, which can be evaluated from
\begin{eqnarray}\label{tauHo}
\left(3^{m-1}\beta A^{2(m-1/2)}\Delta
^{-2q(m-1/2)}\right)|_{t=0}<1,
\end{eqnarray}
in terms of $t_{s}$. In order to satisfy $\tau <H^{-1}$ for all
times we need $m<1/2$. If this condition holds, $\tau$ is a
decreasing function of the cosmic time and becomes equal to zero
at $t_{s}$.  We can take, for example, $\alpha =0$, $m<-1/2$
(from the condition~(\ref{A2}). In other words, with $m<-1/2$
(and $q=1$ and $\alpha =0$), it is possible to have a realistic
models with $\Pi <0$, $|\Pi|<<p$ and $\tau <H^{-1}$ for all times
during the cosmic evolution.

Let us show the behaviour of one of these models, taking for
simplicity, $q=1$, $\alpha =0$ and $m=-1$. In this model the
solution for the Hubble parameter is given by
\begin{eqnarray}\label{Hejemplo}
H=A \Delta ^{-1}.
\end{eqnarray}
From Eq.~(\ref{abcdpunto}) we obtain the expression for the
dissipative pressure
\begin{eqnarray}\label{Piejemplo}
\Pi (\Delta)=- \beta \frac{(A-A_{+})(A-A_{-})}{A^{3}} \Delta ,
\end{eqnarray}
where $A_{\pm}=1/3 (1\pm \sqrt{5})$. As we saw above, $A$ must
satisfy the inequality given in ~(\ref{A}) in order to ensure
$\Pi <0$. The expression for pressure of the GCG in terms of used
Ansatz is
\begin{eqnarray}\label{Pejemplo}
p(\Delta)= -3 \left( \frac{M^{4}}{3}\right)^{(1+\alpha)}
A^{-2\alpha}\Delta ^{2q\alpha},
\end{eqnarray}
which gives for the particular values that we are taking
\begin{eqnarray}\label{Pejemplo1}
p(\Delta)= -M^{4}.
\end{eqnarray}
The negative pressure of the GCG will be constant during the
entire cosmic evolution. The energy density can be obtained from
Eqs.~(\ref{tt}) and~(\ref{Hejemplo}), yielding
\begin{eqnarray}\label{rhoejemplo}
\rho(\Delta)= 3 A^{2}\Delta ^{-2},
\end{eqnarray}
which becomes infinity at $t_{s}$. The upper limit that the
parameter $\beta$ could take can be evaluated from the conditions
$|\Pi|/|p|<<1$ and $\tau <H^{-1}$. Evaluating $|\Pi|/|p|<<1$ for
our model we obtain that
\begin{eqnarray}\label{Pionpejemplo}
\frac{|\Pi (\Delta)|}{|p(\Delta)|}= \frac{\beta}{M^{4}}
\frac{(A-A_{+})(A-A_{-})}{A^{3}} \Delta <<1,
\end{eqnarray}
which implies the following upper limit for $\beta$ when equation
is evaluated at $t=0$
\begin{eqnarray}\label{betalimit}
\beta <<
\frac{3A^{3}}{(A-A_{+})(A-A_{-})}\frac{M^{4}}{3H_{0}^{2}} \left(
H_{0}t_{s}\right)^{-1}.
\end{eqnarray}
The constraint $\tau <H^{-1}$ evaluated at $t=0$ gives the other
restriction on $\beta$
\begin{eqnarray}\label{betalimit1}
\beta < 9H_{0}^{3}.
\end{eqnarray}
If we choose $\beta$ constrained by Eq.~(\ref{betalimit1}),
Eq.~(\ref{betalimit}) can be easily satisfied. This means that
with a enough low viscosity, the dark energy, modeled by a GCG,
leads to a superaccelerated universe.

\section{Characterization of the future-singularities}

We will discuss briefly the conditions which presents the
future-singularities founded in the above sections. In the two
solutions found in section II, for $\alpha >0$, the
future-singularities have the following characterization:

\vspace{0.2 cm}$\bullet$  For $t \rightarrow t_{s}$, $a
\rightarrow \infty$, $\rho \rightarrow \infty$, and $|p|
\rightarrow 0$

\vspace{0.2 cm} The model explicitly calculated for the special
values $q=1, \alpha =0$ and $m=-1$, in section III, gives a
solution which presents a future-singularity characterized by

\vspace{0.2 cm}$\bullet$  For $t \rightarrow t_{s}$, $a
\rightarrow \infty$, $\rho \rightarrow \infty$, and $|p|
\rightarrow $ constant

\vspace{0.2 cm} In the classification realized in~\cite{Odintsov}
exist four types of singularities and none of them behaves like
those founded in this work. In this sense, cosmological models
filled with a dissipative GCG, in the framework of the truncated
Israel-Stewart formalism, present new types of
future-singularities.

\section{Conclusion}

In the present paper we have found cosmological solutions for a
GCG with bulk viscous stresses, in the context of both
thermodynamic formalism of Eckart and Israel-Stewart. We have
assumed that the viscosity has a power-law dependence upon the
density, i.e., $ \xi=\beta \rho^{m}$, where $\beta$ and $m$ are
constant parameters ($\beta \geq 0$). Following the Eckart
approach and in order to obtain solutions, we have derived the
non linear differential equation which leads the evolution of the
Hubble parameter. Analytical solutions are found constraining the
parameters $\alpha$ and $m$ by the relation $m=-(\alpha +1/2)$
and taking particular values of $\beta$. Solutions with
future-singularities can be obtained taking the specific value
$m=1/2$. In this case, the scale factor grows exponentially,
differing from the power law expansion of the scale factor when
the universe is filled with a fluid obeying a barotropic equation
of state and viscosity. As it was discussed in~\cite{Sen},
phantom scenarios with big rip are obtained if $\alpha <-1$ and
$g<0$. Nevertheless, those obtained from models with a GCG and
bulk viscosity avoid the fact that the cosmic fluid could behave
like stiff matter in the past or future evolution of the universe.

Using the causal but reduced version of the Israel-Stewart
formalism we have found a suitable physically solution from the
point of view the thermodynamics requirements expressed in the
conditions $\tau <H^{-1}$ and $|\Pi|/|p|<<1$.  We have explicitly
showed a case in which is possible to cross the barrier $w=-1$.

The future-singularities encounter in our models, in the
framework of the non causal and causal formalism, are of a new
type, not included in the classification presented
in~\cite{Odintsov}.


\section{acknowledgements}
NC and SL acknowledge the hospitality of the Physics Department of
Universidad de La Frontera where part of this work was done. SL
acknowledges the hospitality of the Physics Department of
Universidad de Santiago de Chile.  We acknowledge the partial
support to this research by CONICYT through grant N$^0$ 1040229
(NC and SL). It also was supported from DIUFRO N$^0$ 120618, of
Direcci\'on de Investigaci\'on y Desarrollo, Universidad de La
Frontera (FP) and DI-PUCV, Grants 123.784/06 and 123.105/05 (SL),
Pontificia Universidad Cat\'olica de Valpara\'\i so.

\end{document}